\documentclass[twocolumn,amsmath,amssymb,superscriptaddress,prl,floatfix,footinbib]{revtex4-1}
\usepackage[]{graphicx}
\usepackage{tabularx}
\usepackage[usenames,dvipsnames]{color}
\usepackage{soul}
\usepackage{bm}

\usepackage{times}
\usepackage{amsfonts}
\usepackage{mathrsfs}
\usepackage{graphicx}
\usepackage{dcolumn}
\usepackage{bm}
\usepackage{color}

\usepackage[colorlinks,bookmarks=false,citecolor=blue,linkcolor=red,urlcolor=blue]{hyperref}
\usepackage{multirow}
\usepackage{physics}

\begin{document}

\title{Observation of termination-dependent topological connectivity in a magnetic Weyl kagome-lattice}

\author{Federico Mazzola**}\email{federico.mazzola@unive.it}
\affiliation{Department of Molecular Sciences and Nanosystems, Ca’ Foscari University of Venice, 30172 Venice, Italy}
\affiliation{Istituto Officina dei Materiali, Consiglio Nazionale delle Ricerche, Trieste I-34149, Italy}

\author{Stefan Enzner**}
\affiliation{Institut f\"{u}r Theoretische Physik und Astrophysik and W\"{u}rzburg-Dresden Cluster of Excellence ct.qmat, Universit\"{a}t W\"{u}rzburg, 97074 W\"{u}rzburg, Germany}

\author{Philipp Eck}
\affiliation{Institut f\"{u}r Theoretische Physik und Astrophysik and W\"{u}rzburg-Dresden Cluster of Excellence ct.qmat, Universit\"{a}t W\"{u}rzburg, 97074 W\"{u}rzburg, Germany}

\author{Chiara Bigi}
\affiliation{School of Physics and Astronomy, University of St Andrews, St Andrews KY16 9SS, United Kingdom}

\author{Matteo Jugovac}
\affiliation{Elettra Sincrotrone Trieste S.C.p.A., S. S. 14, km 163.5, 34149 Trieste, Italy}

\author{Iulia Cojocariu}
\affiliation{Elettra Sincrotrone Trieste S.C.p.A., S. S. 14, km 163.5, 34149 Trieste, Italy}

\author{Vitaliy Feyer}
\affiliation{Elettra Sincrotrone Trieste S.C.p.A., S. S. 14, km 163.5, 34149 Trieste, Italy}

\author{Zhixue Shu}
\affiliation{Department of Physics, University of Arizona, Tucson, AZ 85721, USA}

\author{Gian Marco Pierantozzi}
\affiliation{Istituto Officina dei Materiali, Consiglio Nazionale delle Ricerche, Trieste I-34149, Italy}

\author{Alessandro De Vita}
\affiliation{Dipartimento di Fisica Universit\'a di Milano, Via Celoria 16, Milano 20133, Italy}

\author{Pietro Carrara}
\affiliation{Dipartimento di Fisica Universit\'a di Milano, Via Celoria 16, Milano 20133, Italy}

\author{Jun Fujii}
\affiliation{Istituto Officina dei Materiali, Consiglio Nazionale delle Ricerche, Trieste I-34149, Italy}

\author{Phil D. C. King}
\affiliation{School of Physics and Astronomy, University of St Andrews, St Andrews KY16 9SS, United Kingdom}

\author{Giovanni Vinai}
\affiliation{Istituto Officina dei Materiali, Consiglio Nazionale delle Ricerche, Trieste I-34149, Italy}

\author{Pasquale Orgiani}
\affiliation{Istituto Officina dei Materiali, Consiglio Nazionale delle Ricerche, Trieste I-34149, Italy}

\author{Cephise Cacho}
\affiliation{Diamond Light Source, Harwell Campus, Didcot, OX11 0DE, United Kingdom}

\author{Matthew D. Watson}
\affiliation{Diamond Light Source, Harwell Campus, Didcot, OX11 0DE, United Kingdom}

\author{Giorgio Rossi}
\affiliation{Dipartimento di Fisica Universit\'a di Milano, Via Celoria 16, Milano 20133, Italy}

\author{Ivana Vobornik}
\affiliation{Istituto Officina dei Materiali, Consiglio Nazionale delle Ricerche, Trieste I-34149, Italy}

\author{Tai Kong}
\affiliation{Department of Physics, University of Arizona, Tucson, AZ 85721, USA}

\author{Domenico Di Sante}\email{domenico.disante@unibo.it}
\affiliation{Department of Physics and Astronomy, University of Bologna, 40127 Bologna, Italy}
\affiliation{Center for Computational Quantum Physics, Flatiron Institute, 162 5th Avenue, New York, NY 10010, USA}

\author{Giorgio Sangiovanni}\email{sangiovanni@physik.uni-wuerzburg.de}
\affiliation{Institut f\"{u}r Theoretische Physik und Astrophysik and W\"{u}rzburg-Dresden Cluster of Excellence ct.qmat, Universit\"{a}t W\"{u}rzburg, 97074 W\"{u}rzburg, Germany}

\author{Giancarlo Panaccione}\email{panaccione@iom.cnr.it}
\affiliation{Istituto Officina dei Materiali, Consiglio Nazionale delle Ricerche, Trieste I-34149, Italy}

\date{\today}

\begin{abstract}
\bf{Engineering surfaces and interfaces of materials promises great potential in the field of heterostructures and quantum matter designer, with the opportunity of driving new many-body phases that are absent in the bulk compounds. Here, we focus on the magnetic Weyl kagome system Co$_3$Sn$_2$S$_2$ and show how for different sample's terminations the Weyl-points connect also differently, still preserving the bulk-boundary correspondence. Scanning-tunnelling microscopy has suggested such a scenario indirectly. Here, we demonstrate this directly for the fermiology of Co$_3$Sn$_2$S$_2$, by linking it to the system real space surfaces distribution. By a combination of micro-ARPES and first-principles calculations, we measure the energy-momentum spectra and the Fermi surfaces of Co$_3$Sn$_2$S$_2$ for different surface terminations and show the existence of topological features directly depending on the top-layer electronic environment. Our work helps to define a route to control bulk-derived topological properties by means of surface electrostatic potentials, creating a realistic and reliable methodology to use Weyl kagome metals in responsive magnetic spintronics.}

\end{abstract}

\maketitle
** These authors contributed equally

\section{INTRODUCTION}
\begin{figure*}
\centering
\includegraphics[width=1\textwidth]{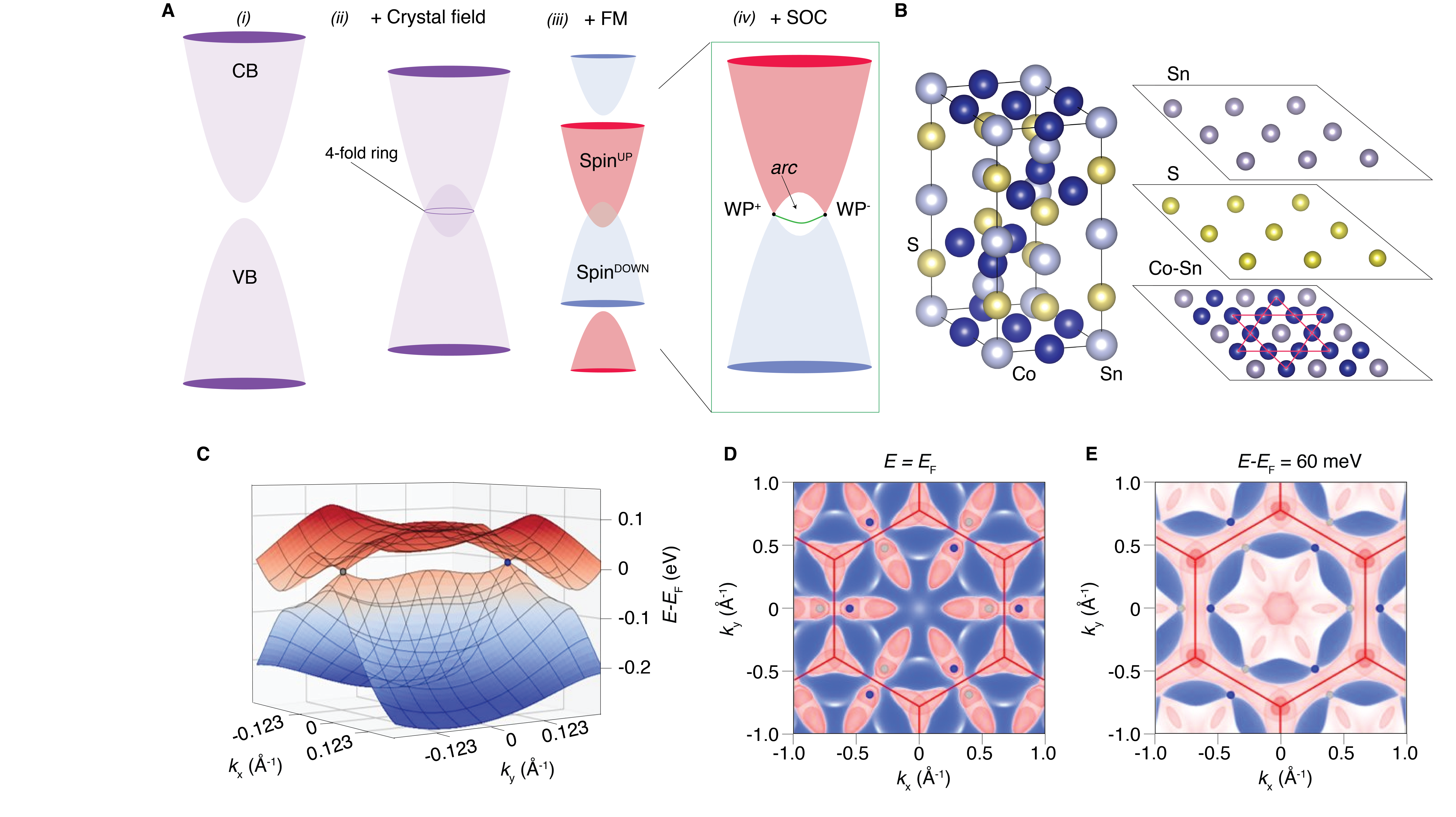}
\caption{(A) Fermi arcs formation mechanism in Co$_3$Sn$_2$S$_2$: (i) The conduction and valence bands (CB and VB) intersect (ii) as a consequence of the crystal-field and form a 4-fold degenerate nodal ring. (iii) The ferromagnetism (FM) splits the bands with opposite spins and the (iv) SOC gaps the electronic structure but two points, the WPs. The WPs are connected by surface states which form arcs in the Fermi surface. (B) Crystal structure of Co$_3$Sn$_2$S$_2$ shown for a single unit cell. The Co-Sn plane have the Co atoms arranged in a kagome lattice; S and Sn form planes with a triangular fashion. (C) Calculated electronic structure around two adjacent Weyl-points. (D) Bulk Fermi surface map obtained with semi-infinite Green's function method corresponding to projections into the $k_\text{z}=0$ plane. (E) Bulk constant energy surface map (60 meV above the Fermi level) of Co$_3$Sn$_2$S$_2$ highlighting the position of the Weyl-points. Increasing spectral intensity is indicated by red, white to blue color.}
\label{Fig1}
\end{figure*}

\begin{figure*}
\centering
\includegraphics[width=0.65\textwidth]{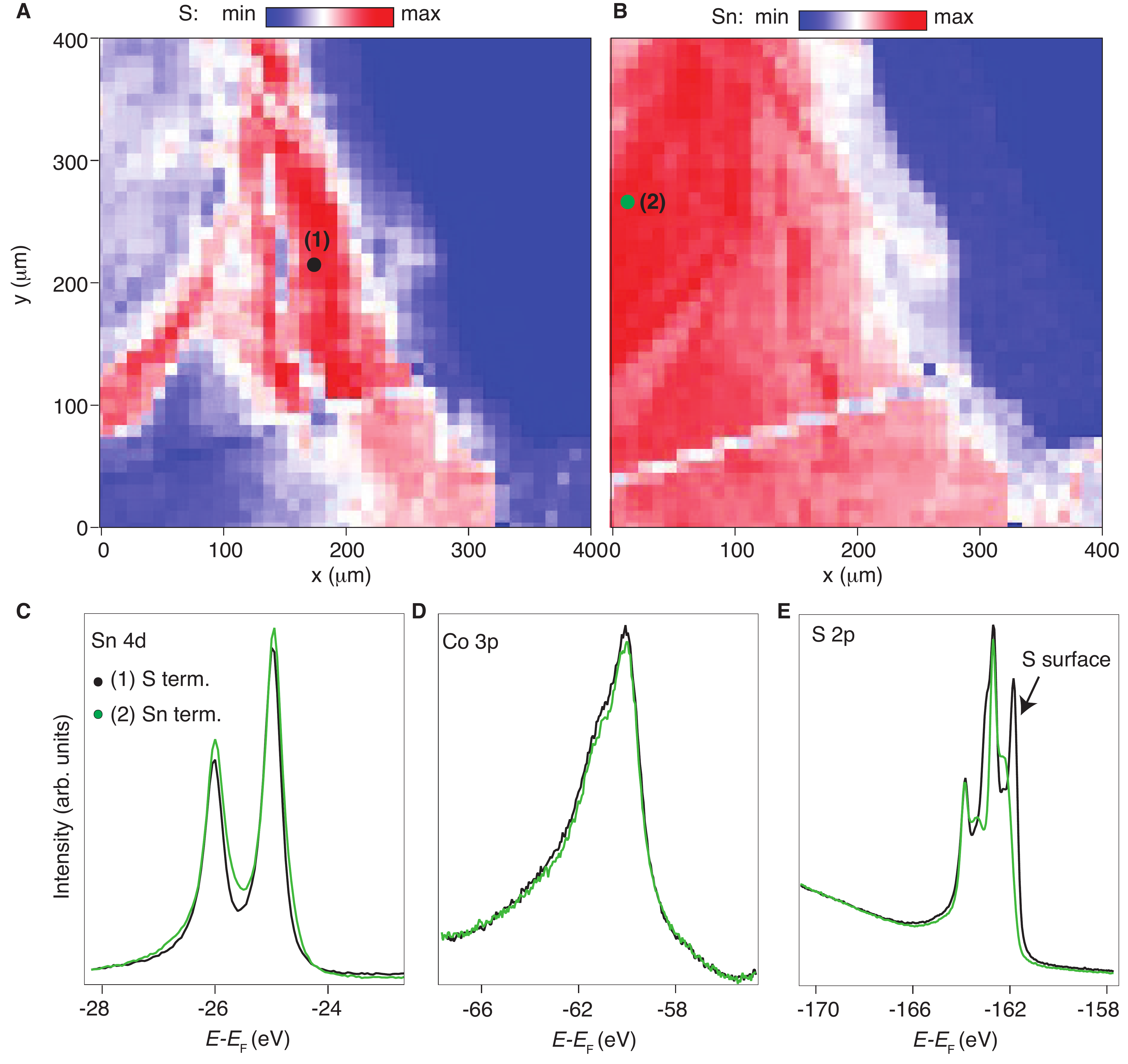}
\caption{(A) Spatial map of a portion of the cleaved sample with red color corresponding to the S termination. (B) The same as (A) but for Sn termination. (C) Sn $4d$, (D) Co $3p$, and (E) S $2p$ core levels taken from points (1) and (2), as indicated in the maps (A) and (B). For the core levels, to better allow us to make a direct comparison, the spectra have been normalized to their tails, i.e. such that for region '1' and '2' each core level has matching tail. In the color maps, the darkest blue regions are areas outside the sample's surface.}
\label{Fig2}
\end{figure*}

The discovery almost two decades ago of oxide heterostructures has given evidence of novel electronic and structural phases which are absent in the bulk compounds, but emerge at the interface between them \cite{Spurgeon_2015, Huang_2018, Sulpizio_2014}. To date, heterojunctions based on transition-metal oxides and dichalcogenides as well as on van-der-Waals systems have been proposed and in some cases also fabricated in the context of spintronics, superconducting devices, solar cells, and low-volatility electronics \cite{Kunstmann_2018, Novoselov_2016}.
In topological quantum materials, the so-called topological protection is believed to be the key for raising interface electronics to a higher level of technological potential. Such developments depend critically on our ability of controlling the properties of topological states at surfaces or interfaces. The bulk-boundary correspondence, a consequence of the non-trivial topology of band structures, implies the existence of protected electronic states at surfaces, edges, and hinges. Yet, their precise arrangement, their electronic connectivity, and the spatial extension of their wave functions are less universal and crucially depend on various materials' properties. For this reason, the knowledge of the surface states in classes of topological compounds are critical for future applications. Magnetic Weyl systems with layered kagome structure belong to this category and their surface electronic properties are a matter of intense study.  Our study provides a solid and comprehensive view on the effect of the surface environment for the electronic properties of Co$_3$Sn$_2$S$_2$.\\

The kagome lattice is a planar network of regular hexagons surrounded by corner-sharing triangles. Materials with a two-dimensional kagome atomic arrangement have been in the spotlight for the study of spin-frustration, arising from the peculiar geometry \cite{Chen_2014,Freedman_2012,You_2012,Meschke_2021,Lachman_2020, Hagemann_2021}. More recently, theoretical predictions, followed by experimental verification, have established kagome lattices as new breed of correlated topological phases, with the mutual existence of high-electron density flat-bands, itinerant graphene-like Dirac states, and the appearance of charge density wave and superconducting phases \cite{Tan_2021,  Ortiz_2020,Kang_2020, Kang_2020C, Kang_2020B, Nakayama_2021, Li_2021}. Additionally, magnetic kagome compounds show unusual transport phenomena, including the anomalous Hall effect and angle \cite{Dhakal_2021, Ma_2021, Tanaka_2020, Min_2022, Li_2019, Shuo_2020}, which are ultimately attributed to the local momentum-enhancement of the Berry curvature \cite{Wang_2018, Okamura_2020, Tanaka_2020, Xu_2018, Liu_2022, Kanagaraj_2022, Liu_2019, Liu_2018}. In time-reversal broken systems, if the spin-orbit coupling (SOC) is sufficiently strong, the orbital mixing can also promote the appearance on non-trivial Chern phases with topological invariants \cite{Fang_2014, Xu_2011}.\\

Of particular interest are magnetic Weyl systems with a kagome structure. In these materials, the combination of crystal field, breaking of time-reversal symmetry (TRS) and SOC is responsible for lifting most of the electronic energy degeneracies in the Brillouin zone (BZ). 
This is not the case for the Weyl-points (WPs) (Fig.\ref{Fig1}a), topologically protected  crossings of Bloch eigenvalues  
in momentum space ($k$-space), which are monopoles of Berry curvature \cite{Ling_2015, Dylan_2020, Chen_2016, Unzelmann_2021}. The latter are responsible for the anomalous Hall effect in several quantum materials \cite{Nagaosa_2010, Taguchi_2001, Haldane_2004, He_2012, Oveshnikov_2015}. In-gap topological surface states connect the WPs and give rise to arc-like features, the so-called Fermi arcs (Fig.\ref{Fig1}a), which participate to the topological properties and transport of Weyl quantum systems. The distribution of the Berry phase in reciprocal space is therefore crucial and the bulk-surface connectivity is tightly bound to it.
In this work, we consider Co$_3$Sn$_2$S$_2$ (Fig.\ref{Fig1}b), which is designed to support SOC and TRS breaking-derived Fermi arcs and has been suggested to host a termination-dependent topological bulk-surface connectivity \cite{Morali_2019, Xu_2018, Xing_2020, Yin_2020, Belopolski_2021}.

\section{RESULTS}

\subsection{Terminations in Co$_3$Sn$_2$S$_2$}

Co$_3$Sn$_2$S$_2$ has three possible cleavage planes, Sn, S, and Co-Sn as shown in Fig.\ref{Fig1}b. Previously, scanning tunnelling microscopy (STM) showed clear evidence for the presence of such multiple surface terminations obtained after cleaving the samples \cite{Morali_2019, Xing_2020}. In addition, the interference pattern of the quasiparticle density of states revealed by the STM tip (so-called quasiparticle interference), demonstrates different scattering channels in momentum space for different terminations \cite{Morali_2019}. This result, supported by density functional theory (DFT) calculations, hints at the possibility that connectivity of the Fermi arcs in Co$_3$Sn$_2$S$_2$, is affected by the local surface environment. Previous photoelectron spectroscopy studies investigated the evolution of this system across the transition temperature \cite{Rossi_2021} and have speculated about the existence of a termination dependent topological connectivity. However, the absence of marked spectroscopic features attributable to the surface in the core levels, makes it difficult to clearly understand the role of the surface in the topological connectivity. Here, we investigate this fundamental issue by studying both the spectra and the Fermi surfaces of Co$_3$Sn$_2$S$_2$ for Sn and S termination and elucidate how the WPs are connected by means of DFT calculations.\\

\begin{figure*}
\centering
\includegraphics[width=0.65\textwidth]{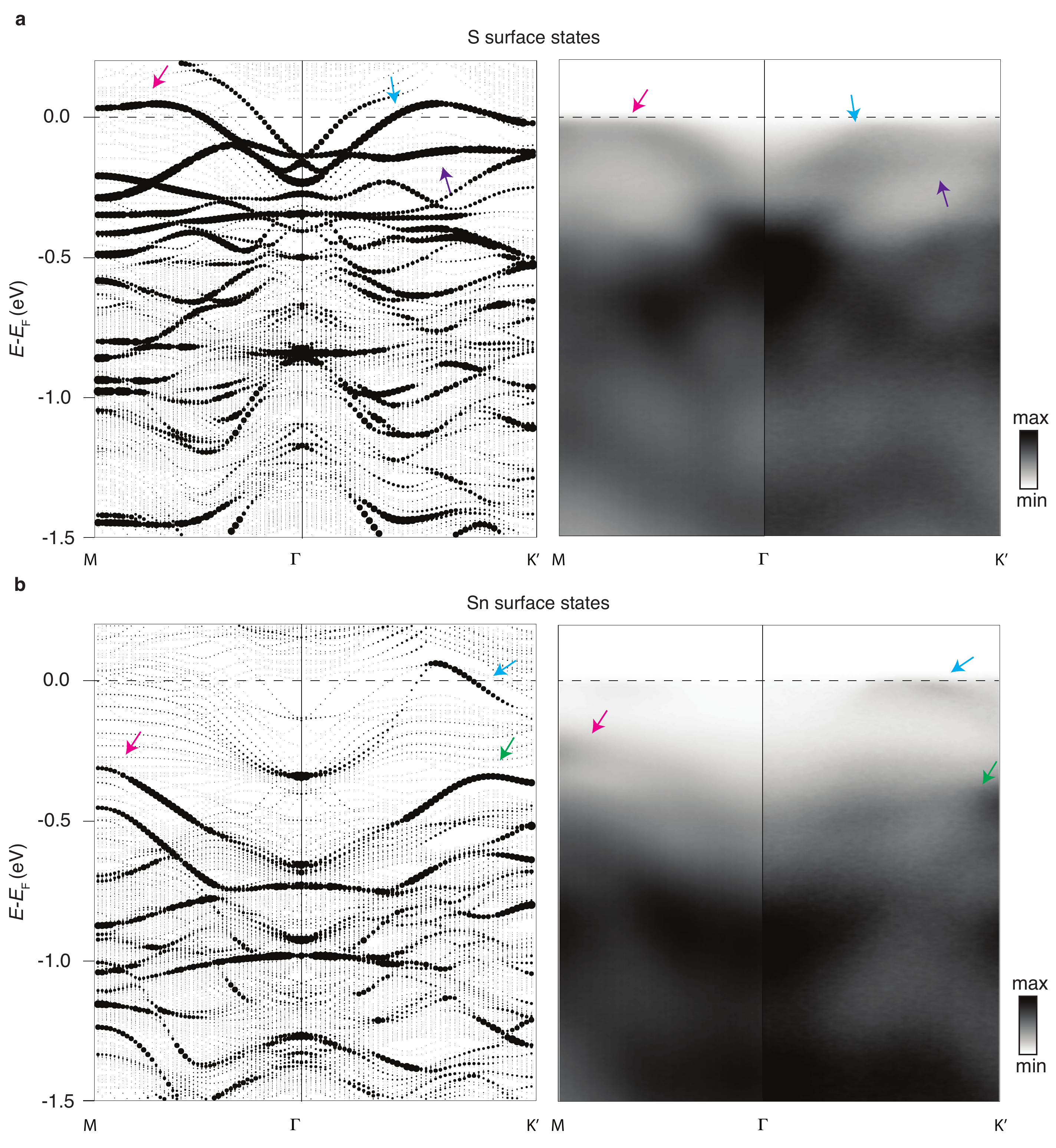}
\caption{(A) DFT calculated energy-momentum spectrum for the S termination (left) and the corresponding ARPES measurement (right) along the high symmetry directions M-$\Gamma$-K$^\prime$. (B) DFT calculated energy-momentum spectrum for the Sn termination (left) and the corresponding ARPES measurement (right) along the high symmetry directions M-$\Gamma$-K$^\prime$. The point size is given by the surface character of the respective termination. The arrows serve to highlight the most prominent common features between the calculations and the ARPES. As one can see, the DFT reproduces extremely well the spectroscopic features seen in ARPES at the Fermi level.}
\label{Fig4}
\end{figure*}

The bulk electronic structure of Co$_3$Sn$_2$S$_2$ through two neighbouring Weyl-points and the bulk Fermi surfaces are shown in Fig.\ref{Fig1}c-e and agree with previously reported works \cite{Liu_2019, Guguchia_2020, Xu_2020}. To examine the role of the surface terminations, we need to characterize this material both in real and reciprocal space. Previous STM measurements report three possible cleavage planes along the (001) direction, giving rise to three distinct situations: Co-Sn, Sn, and S terminated surfaces \cite{Morali_2019} (See Fig.\ref{Fig1}b). The former is reported to have small lateral size of the order of a few nanometers, significantly rarer than the other planes, with the largest concentration of impurities. We will therefore focus on the S and Sn terminations, which could be accessed by photoelectron spectroscopy and provide a suitable platform to study the effect of the surface potential on the bulk-surface connectivity of the WPs. To investigate these terminations, we used micro- and angle-resolved photoelectron spectroscopy (micro-ARPES). The small light spot of micro-ARPES, i.e. 4$\mu$m$^2$ from a capillary mirror, allowed us to probe single/pure termination sample area, thus ensuring the absence of mixed spectroscopic signal. In Fig.\ref{Fig2}a-b, we show the spatially resolved map (each pixel is 4 $\mu$m$^{2}$) where a part of the cleaved sample is measured. The color-code intensity maps denotes where the S  (Fig.\ref{Fig2}a) and the Sn (Fig.\ref{Fig2}b) signals are prominent, respectively (In the scale, red color means more signal, the blue less signal. The plots have been obtained by rastering the same map and selecting the energy corresponding to the core-levels of (a) S and (b) Sn). Our maps are consistent with the STM studies in Ref.\cite{Morali_2019, Xing_2020} and corroborate the previous observation that the Co terminated areas are much smaller than our spatial resolution. From the maps in Fig.\ref{Fig2}a-b we selected two points where the S and Sn intensities were the highest, and we measured accordingly the core levels present in the sample. Consistent with our real space plots, the region '2' in Fig.\ref{Fig2}b shows the highest contribution of the Sn $4d$ core levels (Fig.\ref{Fig2}c), as well as a significantly different shape and reduced contribution for the S $2p$ core levels, compared to the region '1' of Fig.\ref{Fig2}a. The latter shows also a clear surface component shifted towards lower binding energy (Fig.\ref{Fig2}e). \\
This is in contrast to what has been previously reported in Ref.\cite{Rossi_2021}, where there is no evidence of surface-core levels shifts. We stress that a sharp surface-derived signal is crucial to address the surface states manifold, the intensity of which could be otherwise smeared out by aging effects or by the presence of spectroscopic features coming from small neighbouring patches of the sample. The Co $3p$ core levels (Fig.\ref{Fig2}d) showed a much less pronounced difference between the regions. In summary, our micro-ARPES data proves the existence of two main surfaces in Co$_3$Sn$_2$S$_2$ with average $40-50$~$\mu$m$^2$ lateral size for the S and Sn surfaces (Fig.\ref{Fig2}).\\

\subsection{Surface-dependent electronic band structures}

To unveil the differences between the two terminations from the regions '1' and '2' in Fig.\ref{Fig2}, we acquired the energy-momentum spectra by using micro-ARPES along both $\Gamma$-K and $\Gamma$-M directions. In Fig.~\ref{Fig4} we compare these to the DFT calculated spectra (see method sections for experimental conditions and details on the DFT calculations). The availability of micro-ARPES allows us also to make a direct comparison to the data collected by using 'standard' ARPES. The latter are compared for both terminations to the theoretical DFT calculations and are illustrated in the supplementary Fig.S1\,a,c. \\

We emphasize that the theoretical results are obtained in DFT (see Methods) for a supercell of size $65$\,\AA\ in (001) direction, which corresponds to five conventional unit cells. An \textit{ab initio} treatment of the finite size geometry incorporates structural changes and the surface potential correctly, as opposed to non self-consistent slab models based on bulk tight-binding parameters as in Refs.~\cite{Belopolski_2021, Liu_2019, Guguchia_2020, Wang_2018, Xu_2018, Morali_2019}. For the Sn-terminated surface, these effects appear to be of minor importance as both methods yield qualitatively similar band structures and Fermi surfaces (See supplement Fig.~S2). Instead, we find significant differences in the case of S-termination, where the DFT results in better agreement with the experimental data around the Fermi level (see supplement Fig.~S3). \\
The DFT band structure in Fig.~\ref{Fig4} shows the surface states, where the dot sizes indicates the surface localization. According to the DFT, the S termination presents a surface state manifold characterized by a large electron-pocket at the centre of the Brillouin zone and by a smaller pocket with minimum at approximately 100~meV (this feature is more visible in Supplementary figure 1a, for more favorable matrix elements in the data collected by 'standard' ARPES; Here it is present but with significantly weaker intensity). The larger pocket flattens significantly along the $\Gamma$-K and $\Gamma$-M direction approaching the Fermi level. This feature interferes with the other surface states, which were already revealed by the Wannier approach. Since they possess a different degree of surface localization in the DFT calculation a convenient way of highlight the interesting bands giving rise to the Fermi arcs, is a projection onto atoms of the second kagome layer (see supplement Fig.~S4). The Sn termination, in comparison, hosts surface states which cross $E_\text{F}$ away from the zone centre. These are readily revealed by both the DFT and the micro-ARPES in Fig.~\ref{Fig4}.
Contrary to the bands reported in Ref.~\cite{Rossi_2021}, our band structures for the two terminations can clearly be distinguished by various features. This strengthens the identification based on our core level analysis, for which a good agreement between experimental and DFT band structure is observed.\\
We note that the data collected by 'standard' ARPES on of the Sn cleavage (See supplementary Fig.~S1) shows an intense state with electron-like dispersion at the centre of the Brillouin zone. Here, we also notice how such a state matches well with the calculated electronic structure apart from an approximately 100~meV rigid shift of the calculated spectra towards the Fermi level position.\\

In summary, the combination of experiment and theory allows us to reveal a surface-dependent electronic structure, with surface states exhibiting a markedly different dispersion, including both trivial and in-gap topological states. The curvature/flatness of these states is largely influenced by the type of termination, thus the connection of the Weyl-points and the local distribution of the Berry-phase might be also different for the two cases. This observation has important consequences for developing responsive magnetic spintronics, and one can well expect that by applying a surface potential, the topology of the system and the prized effects that derive from it (i.e., anomalous Hall effect, chiral fermions, and Hall angle) can be sensitively tuned for a desired functionality.\\

\subsection{Termination-dependent surface maps and Weyl connectivity}

\begin{figure}
\centering
\includegraphics[width=\columnwidth]{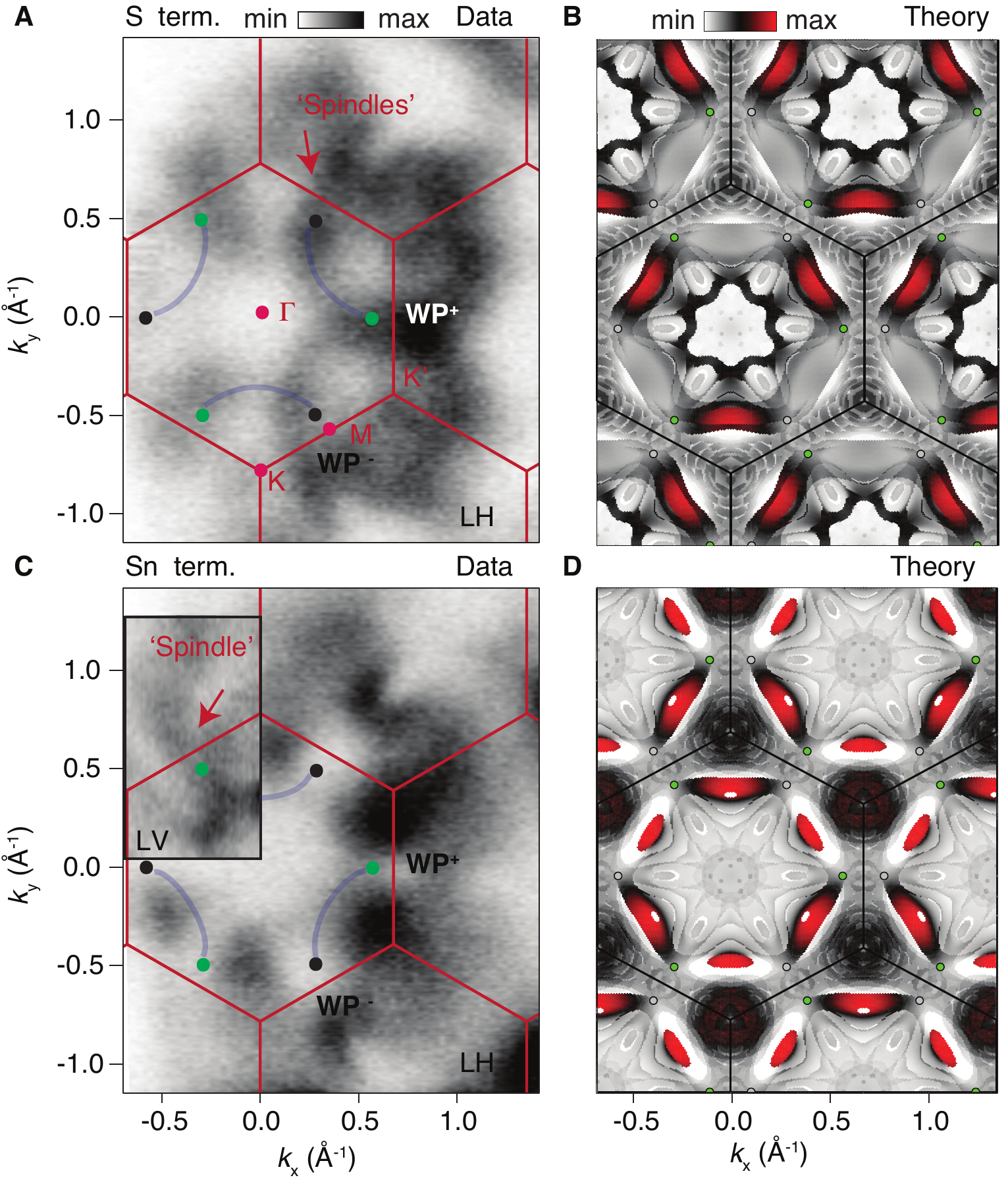}
\caption{(A) Fermi surface map collected at $20$~K by using micro-ARPES from region '1' belonging to the S termination with linear horizontal polarization and (B) corresponding surface map around 70\,meV in a 30\,meV range calculated by DFT with projections onto the second kagome layer. (C) Fermi surface map obtained by micro-ARPES for the Sn termination with linear horizontal polarization and inset showing a detail of the bulk-derived 'spindles' features, obtained by linear vertical polarization. (D) Corresponding DFT S termination at 70\,meV in a 30\,meV range with projections on the first kagome layer. The photon energy used was the same, i.e., 120 eV. Fermi surfaces with linear vertical polarization are available in the supplementary information. The Weyl-points, are also reported together with the expected arcs features.}
\label{Fig3}
\end{figure}

The Fermi surfaces of Co$_3$Sn$_2$S$_2$ for the S- and Sn-terminations have been collected with micro-ARPES in the same experimental conditions and they have been compared to the \textit{ab initio} Fermi surfaces calculated by using DFT. In Fig.\ref{Fig3} we show both experimental and calculated results. We stress that the three-dimensional nature of Co$_3$Sn$_2$S$_2$ is responsible for broadening the electronic structure and causes strong variations in the photoemission intensity when varying photon energies. Thus, our combined approach, involving both theory and experiment, is crucial to identify the main components of the electronic structure. \\
Co$_3$Sn$_2$S$_2$ exhibits elongated features along the $\Gamma$-M direction, bridging neighbouring Brillouin zones. We will refer to them as 'spindles' \cite{Kanagaraj_2022, Wang_2018}. These characteristics are bulk properties and can already be observed for the bulk Fermi surface in Fig.\ref{Fig1}d. The spindles are also noticeable in the slab calculations (Fig.\ref{Fig3}b,d).
Experimentally we detect the spindles at both terminations, yet for different intensity depending on the light polarization. For linear horizontal polarization, these spectroscopic features are the most intense at the S-terminated surface (Fig.\ref{Fig3}a). On the contrary, we do not see spindles for the Sn-termination with the same light (Fig.\ref{Fig3}c). However, they are recovered by using linear vertical light, as visible in the inset of Fig.\ref{Fig3}c, where a portion of the Fermi surface in such a favorable condition has been illustrated (See also supplementary Fig.~S5). Already the difference with regards to bulk like features further proves the presence of the two distinct termination Sn and S, as identified by the core level analysis.
\\

Next we focus on characteristics originating from the surface states. The DFT surface maps shown in Fig.\ref{Fig3}b,d are obtained at an energy close to the Weyl-points to identify their connectivity. The S-terminated surface map is dominated by the presence of trivial surface states of the large pocket. Projecting however on the second Co-kagome layer, we strongly suppress such trivial states, highlighting the actual Weyl connectivity around the K point in Fig. \ref{Fig3}b. The projection on the first surface layer can be seen in Fig. S12B.
The linear vertical polarization is a good condition for observing surface states in this case (See supplementary Fig.S5 A,LV). The combination of trivial, non-trivial surface states and spindles results into an overall hexagonal pattern in the middle of the BZ (Fig.S5 A,LV). Experimentally, it is difficult to detect all the theoretically predicted surface states. This is not only challenged by the strongly varying matrix elements (See supplementary information) and the complexity of the electronic structure of this system itself, but also by the presence of bulk bands (due to their large $k_\text{z}$ dispersion). These appear very broad and often make the identification of the surface states (which are generally sharper in nature) difficult to be seen. The identification of the main features in Fig.\ref{Fig3}a-b is possible thanks to comparison with DFT. However, the complexity of the system and the strongly varying matrix elements make it very challenging to understand by ARPES the subtle differences observed from K and K$^\prime$. In the supplementaty information (Fig. S8), we have also shown additional data collected closer to the photon energy corresponding to the Weyl-point and we could, in such a configuration, identify a three-fold symmetric pattern, like the one predicted by the DFT. The latter allows us to identify the spindles and the intra-BZ connections of the Weyl-points with Fermi arcs around the K points (and not the K$^\prime$).\\
In contrast, the Sn termination shows a Fermi surface with spectroscopic features mainly located around the K- and K$^\prime$-points of the BZ for linear horizontal polarization. Such characteristics are expected based on the triangular surface state from DFT and contribute to the spectral weight with the largest signal (Fig.\ref{Fig3}c). Additionally for the Sn-termination, our DFT shows arcs which connect the WPs in 'opposite fashion' compared to the S-termination, winding around the K$^\prime$ points (and not the K). The pairing can also be obtained from the surface band structure through adjacent Weyl-points, as shown in supplementary Fig. S11, where we characterize non-trivial surface states linking the Weyl-points. For better identification of the Weyl-point connection we utilized a rather large energy interval of 30\,meV for the surface maps. In the supplementary Fig. S12, we depict typical Fermi arcs on surface maps by using a smaller energy range. Also in the Sn-terminated case, experimentally we see an overall hexagonal pattern with intensity connecting the WPs, due to the combination of trivial surface states and Fermi arcs (See supplementary Fig.S5 B,LV).\\
It is worth noticing that the Fermi surfaces have been purposely collected and shown in the same conditions to better appreciate the strong spectroscopic differences observed. These differences show a distinct fermiology between the S- and the Sn-terminated samples. The experimental Fermi surfaces show both trivial and non-trivial surface states which connect the Weyl-points. Even though by changing the photon energy, it is possible to enhance the intensity of the surface states, resulting in a simpler identification of these (See supplementary Figure S5), and despite the one-to-one comparison between calculations and ARPES is overall good, it is still challenging to distinguish between trivial and non-trivial surface states. However, the evident differences observed in our data are still important to show how the surface electronic enviroment is crucial to modulate both topological and topologically trivial components. We show this in the supplementary information for completeness. The data shown here are collected at one photon energy for which the spectral features from both terminations were identified with comparable intensity.\\

In summary, the two surface cleavage planes give a markedly different Fermi surface for Co$_3$Sn$_2$S$_2$. We emphasize that the importance of these findings has a direct link to the development of highly controllable device schemes, where the use of electrostatic potential, in a manner akind field effect transistors, can be used to finely modulate the topology and in particular its local character. We additionally mention that experimentally and theoretically the Sn termination is in agreement with previous works (See 'standard' ARPES carried out at lower photon energy, i.e., 55-65 eV in supplementary Fig.~S7, which allow us to get the details of the Fermi surface around the K- and K$^\prime$ points, revealing Fermi arcs similar to those of Ref.\cite{Liu_2019, Belopolski_2021}). In this study, the great advantage of micro-ARPES is the possibility of getting simultaneously lateral dimensions, core-level, and ARPES information of a single sample's termination. In addition, we stress that this was done for both S and Sn terminations by using the same experimental conditions, i.e., geometry, photon energy, and both light polarizations. This is a necessary requirement for an unambiguous differentiation of the S and the Sn surface, as the spectroscopic signal uniquely from a single termination. In this sense, our approach allows us to prove unequivocally that the Sn and S terminations host significantly different Fermi surfaces with distinctive connectivity. Depending on the termination the Fermi arcs connect different Weyl-points inside the BZ.\\
Our results proves that the surface environment is crucial for determining the shape of the arcs connecting the WPs, thus revealing that the bulk-surface connectivity is strongly altered by the type of termination. This situation is similar to what happens in polar materials \cite{Mazzola_2018, Mazzola_2022}, however, in the present case it has implications in the k-resolved local topology, which is determined by the fashion that the WPs are connected, as shown in Ref.\cite{Morali_2019}. Therefore, in this particular system, the large differences observed by our experiment are important to pave the way for the future perspective of using electric fields to tune the transport and topological properties of Co$_3$Sn$_2$S$_2$.\\

\section{DISCUSSION}

In conclusion, by using micro-ARPES and DFT calculations, we not only verified the existence of two micrometer size terminations on Co$_3$Sn$_2$S$_2$, but we also unambiguously demonstrate how in this compound the same bulk topology gives rise to striking differences of the Weyl-point connectivity on each of the two surface terminations. To prove this, it was crucial to collect both core levels, Fermi surfaces, and energy-momentum spectra from the same micrometric and single domain surface area. In Co$_3$Sn$_2$S$_2$, we show visibly different topological surface states dispersing across the Fermi energy. Here, we shed light on the possibility to modulate the surface electronic environment of Co$_3$Sn$_2$S$_2$ by using electrostatic potentials to shape the transport properties of this system, \textit{in-primis} dictated by the interplay of SOC and magnetism. Potential future implications of this study are expected to emerge in the field of designer heterostructures design.\\


\noindent{\bf Methods}\\

The samples were grown by a self-flux method \cite{Liu_2019}. Starting elements were loaded in an alumina crucible in a stoichiometric ratio and sealed in a silica tube under vacuum. The silica ampule was then heated to 1000~°C, kept there for 24 hours and slowly cooled to room temperature. Micro-ARPES measurements were performed at the I05 beamline at Diamond Light Source, UK. A capillary mirror optic was used to focus the beam to $\sim$4 $\mu$m, and a DA30 hemispherical analyser was used to measure the photoelectrons with a combined energy resolution of ~30 meV. Fermi surface maps were collected by rotating the analyser around the fixed sample position. The 'standard' ARPES has been performed at the APE-LE laboratory of NFFA facility of IOM-CNR at the Elettra synchrotron radiation source (Trieste, IT - https://www.nffa.eu) by using a DA30 hemispherical analyser and at the laboratory nanoESCA of the Elettra synchrotron. All the photoemission data have been collected at $\approx 20$~K, temperature well below the sample's ferromagnetic transition (T$_\text{C}$=175 K) and at the base pressure of $1\times 10^{-10}$~mbar\\

The DFT calculations of magnetic Co$_3$Sn$_2$S$_2$ were performed as implemented in the Vienna \textit{ab initio} simulation package (VASP) \cite{VASP}, within the projector augmented-plane-wave (PAW) method \cite{PAW1,PAW2}. For the exchange-correlation potential the PBE-GGA functional~\cite{PBE} was used, by expanding the Kohn-Sham wave functions into plane waves up to an energy cutoff of $400\ $eV. We sample the Brillouin zone on a $8\times8\times1$ Monkhorst-Pack mesh and employ spin-orbit coupling self-consistently \cite{SOC_VASP}.\\
To investigate the electronic structure of the surface we calculate a slab geometry of five conventional unit cells in z-direction with a vacuum distance of 15\,\AA. All atoms in the outer layers are allowed to relax in non-periodic direction until forces were below 0.01\,eV/\AA. At the S-termination the magnetic moments for the Co-atoms were constrained to the magnetic moments in the bulk region. See supplementary for more details.\\
For the bulk Fermi surfaces in Fig.1, we project the Bloch wave functions of the conventional bulk cell onto maximally localized Wannier functions based on Co $3d$, Sn $5p$ and S $4p$ orbitals and construct a tight-binding Hamiltonian using the WANNIER90 package \cite{wannier}. We calculate bulk Fermi surfaces using the Green's function method with WannierTools \cite{wanniertools}. The Fermi surfaces shown in Fig.\ref{Fig3} where done with DFT in VASP. Solely for comparison to our supercell DFT approach, we obtain the band structure of a Wannier slab. (See supplementary Fig.S2, Fig.S3 and Fig.S4).


\bibliographystyle{naturemag}
\bibliography{biblio}

\ \\

\noindent{\bf Acknowledgements}\\
The research leading to these results has received funding from the European Union's Horizon 2020 research and innovation program under the Marie Sk{\l}odowska-Curie Grant Agreement No. 897276. We gratefully acknowledge the Gauss Centre for Supercomputing e.V. (https://www.gauss-centre.eu) for funding this project by providing computing time on the GCS Supercomputer SuperMUC-NG at Leibniz Supercomputing Centre (https://www.lrz.de). We are grateful for funding support from the Deutsche Forschungsgemeinschaft (DFG, German Research Foundation) under Germany's Excellence Strategy through the W\"urzburg-Dresden Cluster of Excellence on Complexity and Topology in Quantum Matter ct.qmat (EXC 2147, Project ID 390858490), through FOR 5249-449872909 (Project P5) as well as through the Collaborative Research Center SFB 1170 ToCoTronics (Project ID 258499086). We greatly acknowledge Diamond light Source who supported the entire micro-ARPES experiment and corresponding costs. The Flatiron Institute is a division of the Simons Foundation. PK and CB gratefully acknowledge support from The Leverhulme Trust via Grant No. RL-2016-006\\


\noindent{\bf Competing financial interests}\\
the Authors declare no Competing Financial or Non-Financial Interests

\clearpage
\newpage

\section{Supplementary Information}

\section{Micro-ARPES spatial maps plotting}

In order to plot the real space maps for the S and Sn terminations shown in Fig.2a-b of the main text, we did the following: We colored the maps by selecting the energies of the S (and Sn) core levels for the full area. In particular, (1) For the S termination, we identified features in the micro-ARPES spectra which were present only when the surface component of the S $2p$ core levels was visible. Then, we tracked the regions across the probed area, where such features showed up. (2) for the Sn termination, we used the same idea and we identified spectroscopic feature which were only visible when the signal of the Sn $4d$ core levels were the highest and the surface component of the S $2p$ core levels was absent. It is also worth mentioning that we have collected the same maps for Co core levels and we did observe a constant intensity except a few small areas which, however, did not gove any coherence in the ARPES spectra, likely indicating the disordered character of this surface compared to S and Sn.

\section{DFT details}
The magnetic moments of the cobalt atoms were 0.338 $\mu_\text{B}$ in the bulk DFT calculation, which was used for the Wannier model. The tin atoms showed small magnetic moments on the order of -0.020$\mu_\text{B}$ to -0.030$\mu_\text{B}$. The magnetization was exclusively perpendicular to the kagome layers. \\
In the DFT supercell slab calculation of five conventional unit cell in the z-direction, the magnetic moments of the cobalt atoms in the bulk region were found to be 0.34$\mu_\text{B}$ to 0.35$\mu_\text{B}$. The S-terminated surface exhibit significantly reduced magnetic moments, however they were constrained to the values of the bulk region.

\section{Matrix elements, photon energy, and three-fold spectral weight}
In this section we show additional experimental data which allow us to better understand the role of matrix elements in the electronic structure, dimensionality of the observed bands, and also help the visualization of certain features. Indeed, matrix elements are of paramount importance in photoemission experiments and can determine the intensity of the ARPES maps significantly. Such matrix elements depend on the photon energy, photon polarization, and also on the geometry of the system. In order to shed light on this aspect, here, we report the study of the electronic properties of Co$_3$Sn$_2$S$_2$ with different experimental configurations. In Fig.\ref{S1}, the Fermi surface maps of the system are reported from both (a) S and (b) Sn terminations from areas of 4$\mu$m$^2$ and for both linear vertical and linear horizontal configurations. Such configurations correspond to having the light polarization vector fully in-plane and (parallel to the analyzer slit) and also 50$\%$ in-plane (orthogonal to the analyzer slit) and 50$\%$ out of plane. We notice that the data highlights a strong linear dichroism, i.e. different polarizations yield very different spectroscopic signals. Importantly, by using both polarizations we can detect the "spindles", which appear for both terminations (as in our calculations) along the $\Gamma$-M direction of the Brillouin zone. This observation is particularly important, because it allows us to exclude the presence of rotational domains, leading to the wrong attribution of the spectral weight measured. In other words, if the observed features were given by domains, the "spindles", in one termination case, would not be along $\Gamma$-M. To better visualize this, we also show for both surfaces, both light polarizations, and both $\Gamma$-K and $\Gamma$-M directions the electronic structure as energy versus momentum and we found strikingly different dispersions between S and Sn terminated samples. Finally, we also have reported the Fermi surface maps by rotating the sample by 90 degrees, and we have made sure that the features observed in Fig.\ref{S1} and \ref{S4} are consistent with our attribution. We reported this in Fig.\ref{rot}.

In order to give a comprehensive understanding of the electronic properties of this system, we also show in the following photon energy dependent scans collected for one termination (Sn-termination) showing the two-dimensional character of the states observed. Also, as an example, we selected an energy close to the Weyl points (137 eV) at which the three-fold symmetric pattern expected for a system where K and K' are not equivalent, shows up better and where the features that we compare to the DFT in the main text are better visible (See Fig.\ref{ekz}).

\clearpage
\newpage

\begin{figure*}[!h]
\centering
\includegraphics[width=0.6\textwidth]{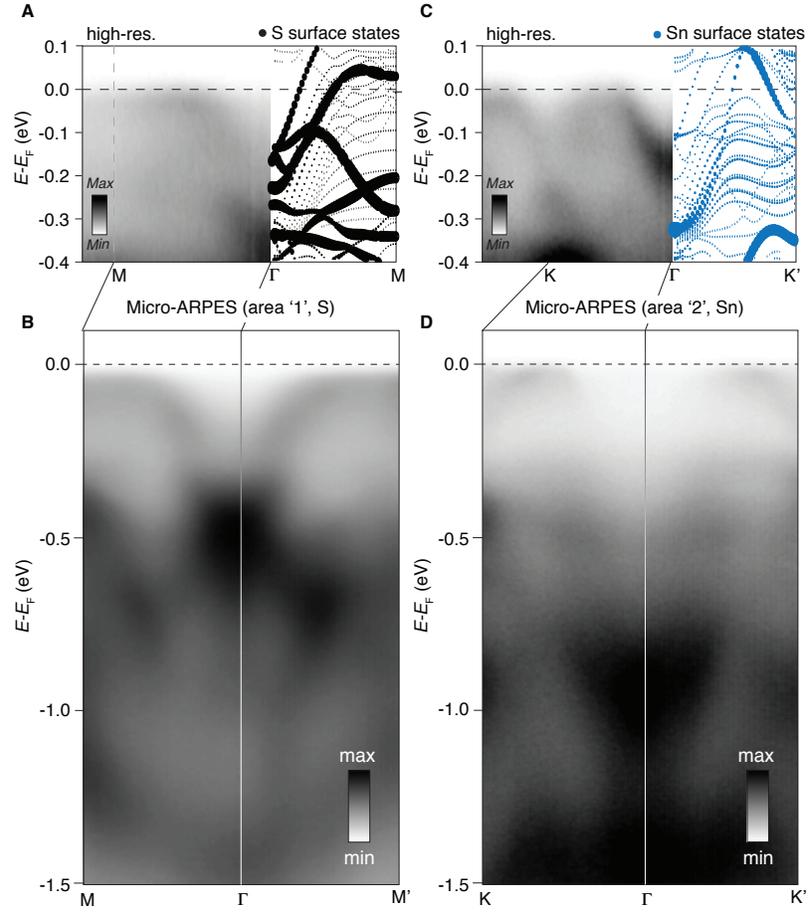}
\caption{(A) High-resolution ARPES along the $\Gamma$-M direction and corresponding DFT calculated spectrum obtained for the S surface states. (B) Extended range obtained from spot '1' by using micro-ARPES.(C) High-resolution ARPES along the $\Gamma$-K$^\prime$ direction and corresponding DFT calculated spectrum obtained for the Sn surface states. (D) Extended range obtained from spot '2' by using micro-ARPES. The measurements have been performed at at 20 K. We notice that the high-resolution data also show an electron-like pocket at the centre of the zone which is reconcilable with the DFT calculations with a simple 100 meV rigid shift towards E$_\text{F}$.}
\label{S5}
\end{figure*}

\begin{figure*}[!h]
\centering
\includegraphics[width=1\textwidth]{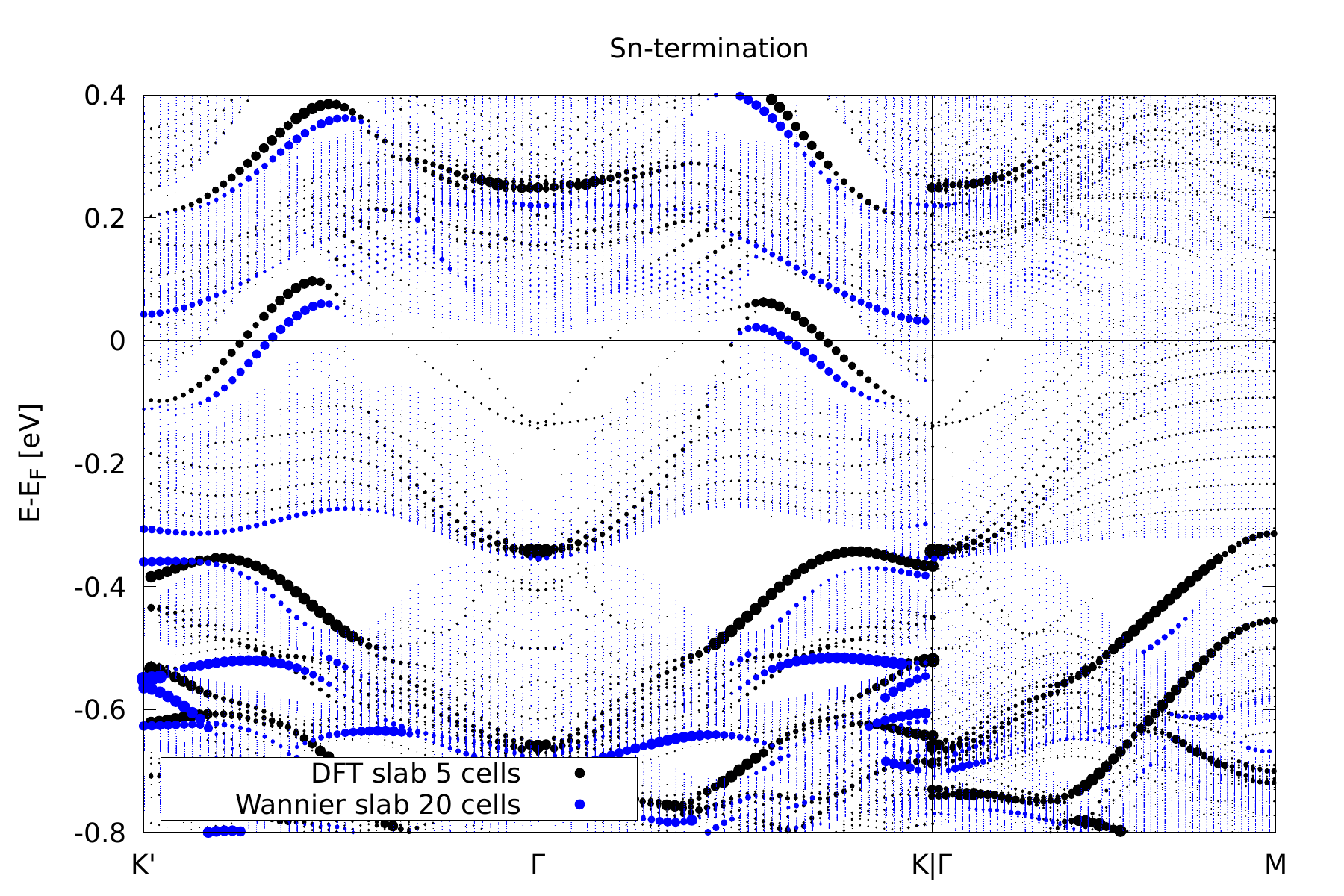}
\caption{Comparison of surface states obtained with different methods for the Sn-termination along the K$^\prime$-$\Gamma$-K$\mid$$\Gamma$-M path. The surface band structure for a supercell, consisting of five conventional cells stacked in z-direction, as calculated with DFT is shown in black. The corresponding band structure from a Wannier Hamiltonian based on the conventional unit cell is depicted in blue. The size of the dots indicates the localization in the first three layers. Differences in the Fermi energies between the two methods may arise, as the Fermi energy is not calculated self-consistently for the Wannier slab approach.}
\label{S6}
\end{figure*}

\begin{figure*}[!h]
\centering
\includegraphics[width=1\textwidth]{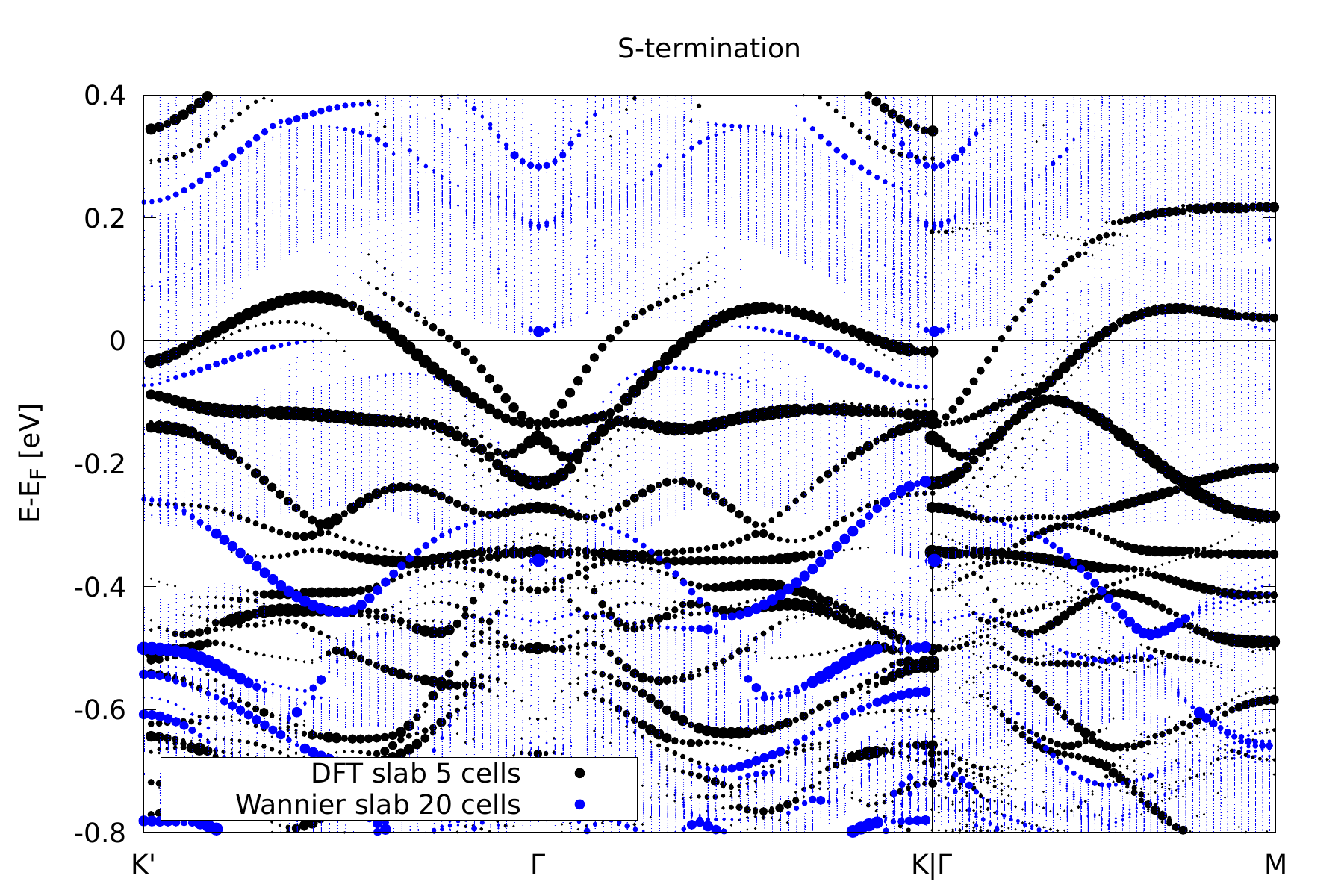}
\caption{Comparison of surface states obtained with different methods for the S-termination along the K$^\prime$-$\Gamma$-K$\mid$$\Gamma$-M path. The surface band structure for a supercell, consisting of five conventional cells stacked in z-direction, as calculated with DFT is shown in black. The corresponding band structure from a Wannier Hamiltonian based on the conventional unit cell is depicted in blue. The size of the dots indicates the localization in the first two layers. In the DFT approach we observe additional surface states not captured by the Wannier approach i.e. the pocket around the Gamma point, which becomes flat towards the K and M point.}
\label{S7}
\end{figure*}

\begin{figure*}[!h]
\centering
\includegraphics[width=1\textwidth]{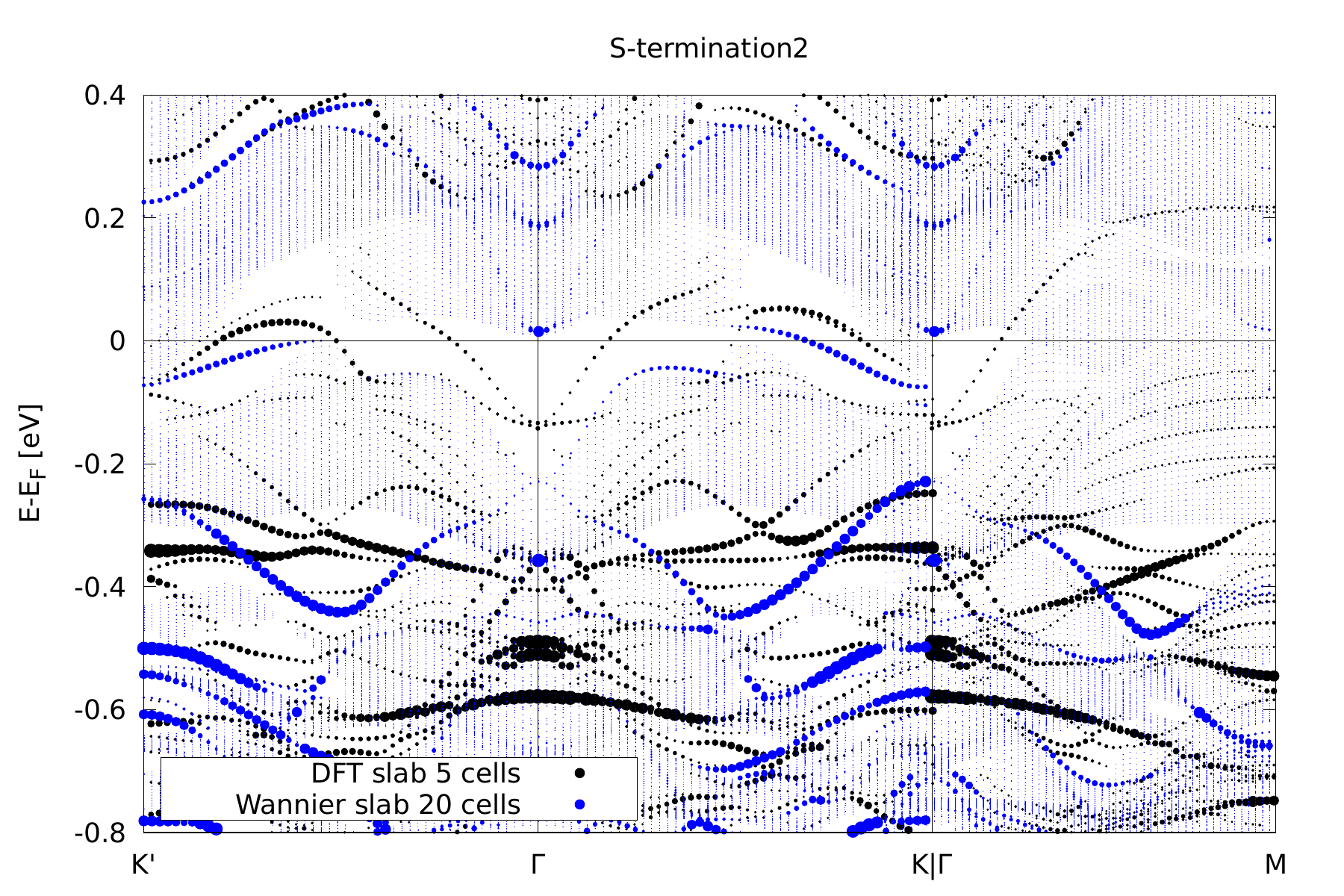}
\caption{Comparison of surface states obtained with different methods for projections of the second outermost Co-kagome layer of the S-termination along the K$^\prime$-$\Gamma$-K$\mid$$\Gamma$-M path. The surface band structure for a supercell, consisting of five conventional cells stacked in z-direction, as calculated with DFT is shown in black. The corresponding band structure from a Wannier Hamiltonian based on the conventional unit cell is depicted in blue. The size of the dots indicates the localization in the second kagome layer including adjacent S-layers. Here we exploit that the surface states, already noticed from the Wannier calculation, also localize on second Co-kagome layer, whereas the additional surface states revealed by DFT are more restricted to the first kagome layer.}
\label{S7}
\end{figure*}

\begin{figure*}
\centering
\includegraphics[width=0.5\textwidth]{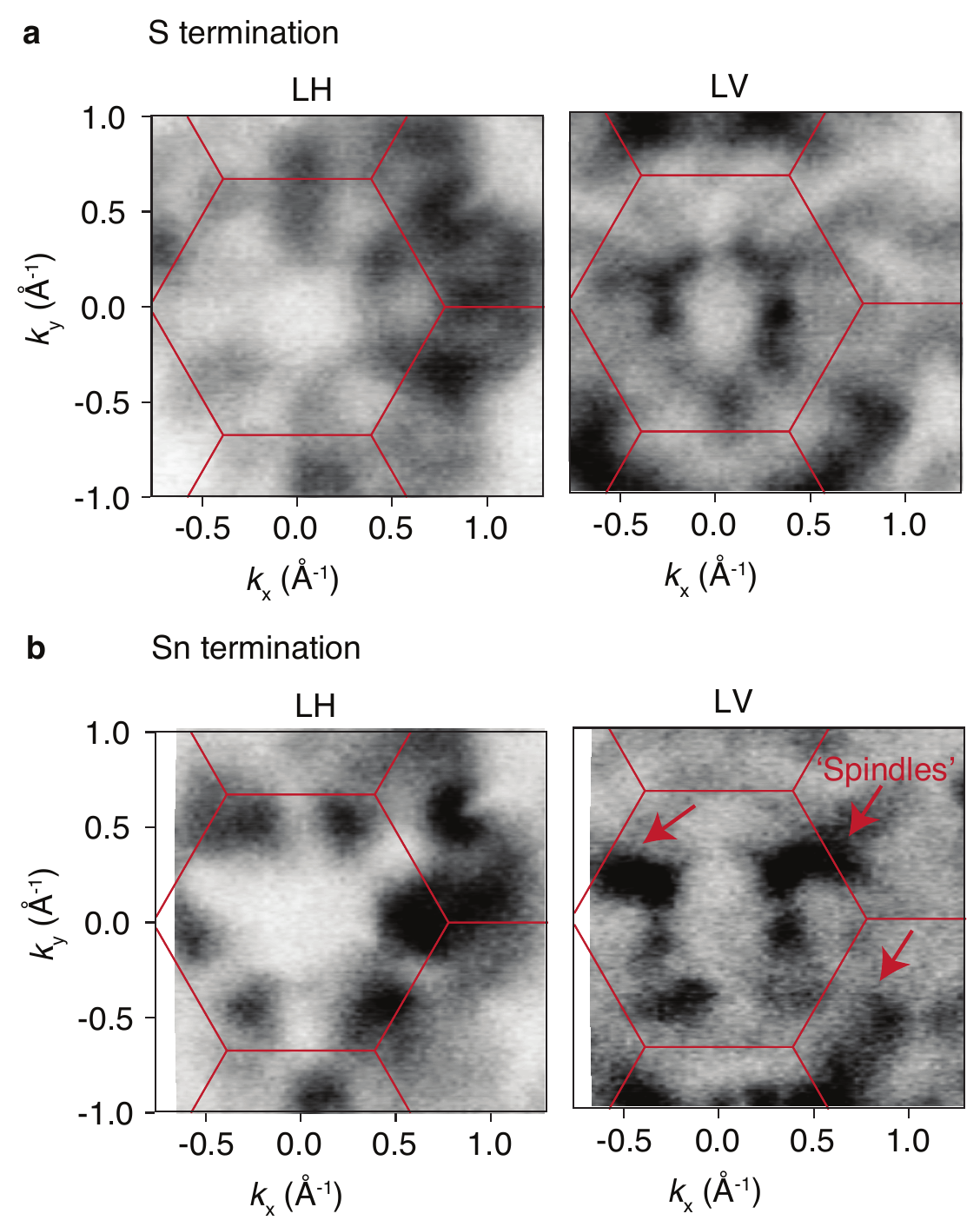}
\caption{(a) S termination Fermi surface collected with linear horizontal (left) and vertical (right) light. (b) Sn termination Fermi surface collected with linear horizontal (left) and vertical (right) light. }
\label{S1}
\end{figure*}

\begin{figure*}
\centering
\includegraphics[width=\textwidth]{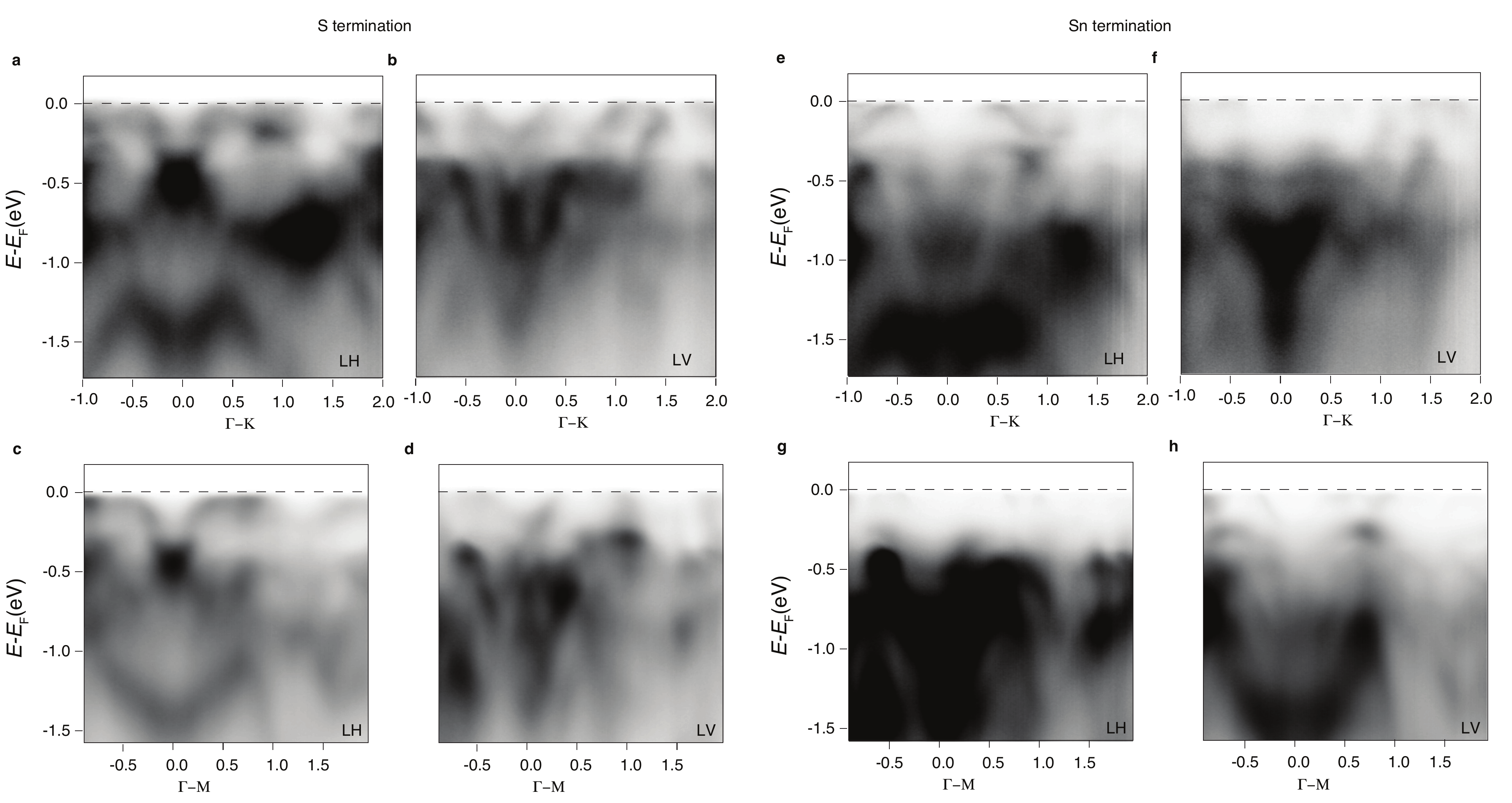}
\caption{S termination collected at 120 eV with (a) linear horizontal and (b) linear vertical polarization along the and along the $\Gamma$-K direction. The same collected with (c) linear horizontal and (d) linear vertical polarization along the and along the $\Gamma$-M direction. Sn termination collected at 120 eV with (e) linear horizontal and (f) linear vertical polarization along the and along the $\Gamma$-K direction. The same collected with (g) linear horizontal and (h) linear vertical polarization along the and along the $\Gamma$-M direction.}
\label{S4}
\end{figure*}

\begin{figure*}
\centering
\includegraphics[width=0.5\textwidth]{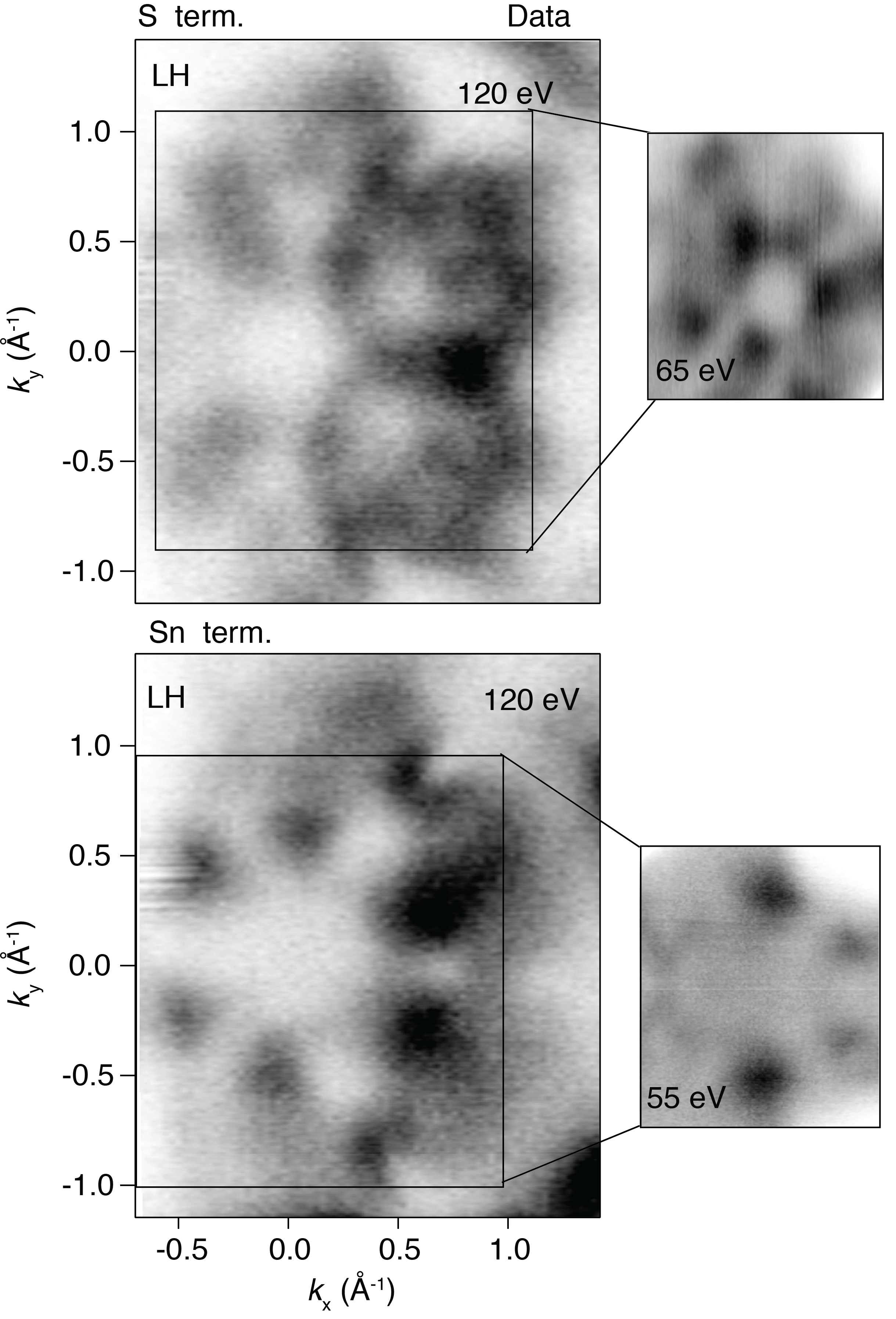}
\caption{Termination-dependent Fermi surface maps collected with linear horizontal light after the sample has been rotated by 90 degrees compared to Fig.\ref{S1}. We notice that for both terminations, the spectral features remain the same.}
\label{rot}
\end{figure*}

\begin{figure*}
\centering
\includegraphics[width=\textwidth]{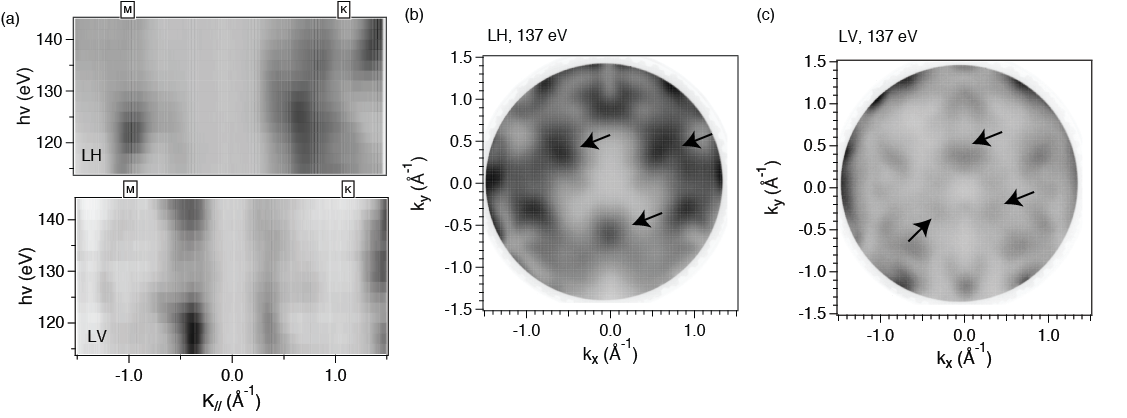}
\caption{(a) Photon energy scan collected for the Sn-termination with both LH and LV polarization. The stripe-like behaviour is in agreement with previous works and indicates the two-dimensional character of the electronic states. Selected photon energy (137 eV) close to the Weyl points for (b) LH and (c) LV polarization where the three-fold pattern (due to K and K' difference) is more prominent. The three-fold symmetric intensity is also indicated by the arrows.}
\label{ekz}
\end{figure*}

\begin{figure*}
\centering
\includegraphics[width=0.5\textwidth]{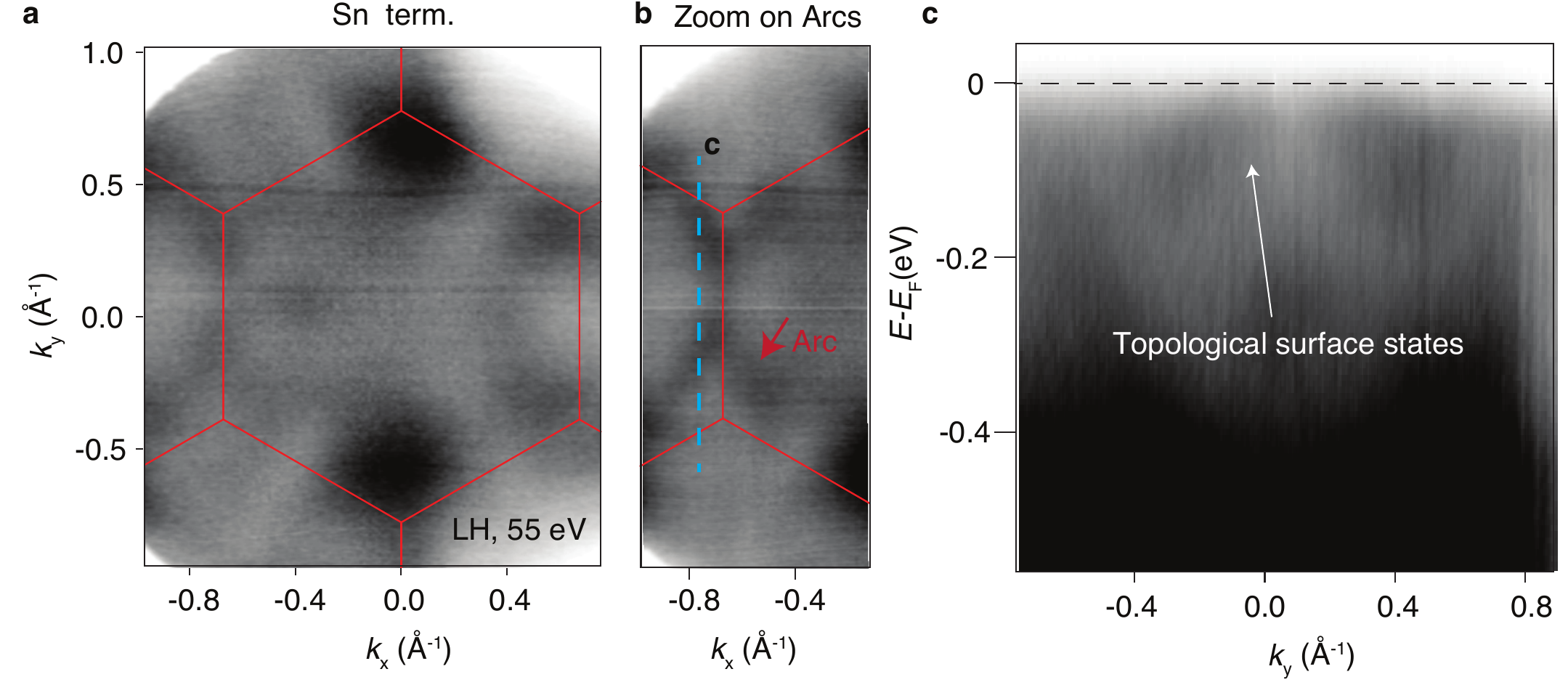}
\caption{(a) Sn termination Fermi surface collected with linear horizontal polarization and at 55 eV and its corresponding (b) zoom on the Fermi arcs. Such arcs form a triangular patter around the K point of the Brillouin zone. (c) Energy-momentum spectra extracted along the blue dashed line in (b) and showing the topological surface states.}
\label{S2}
\end{figure*}

\begin{figure*}
\centering
\includegraphics[width=0.5\textwidth]{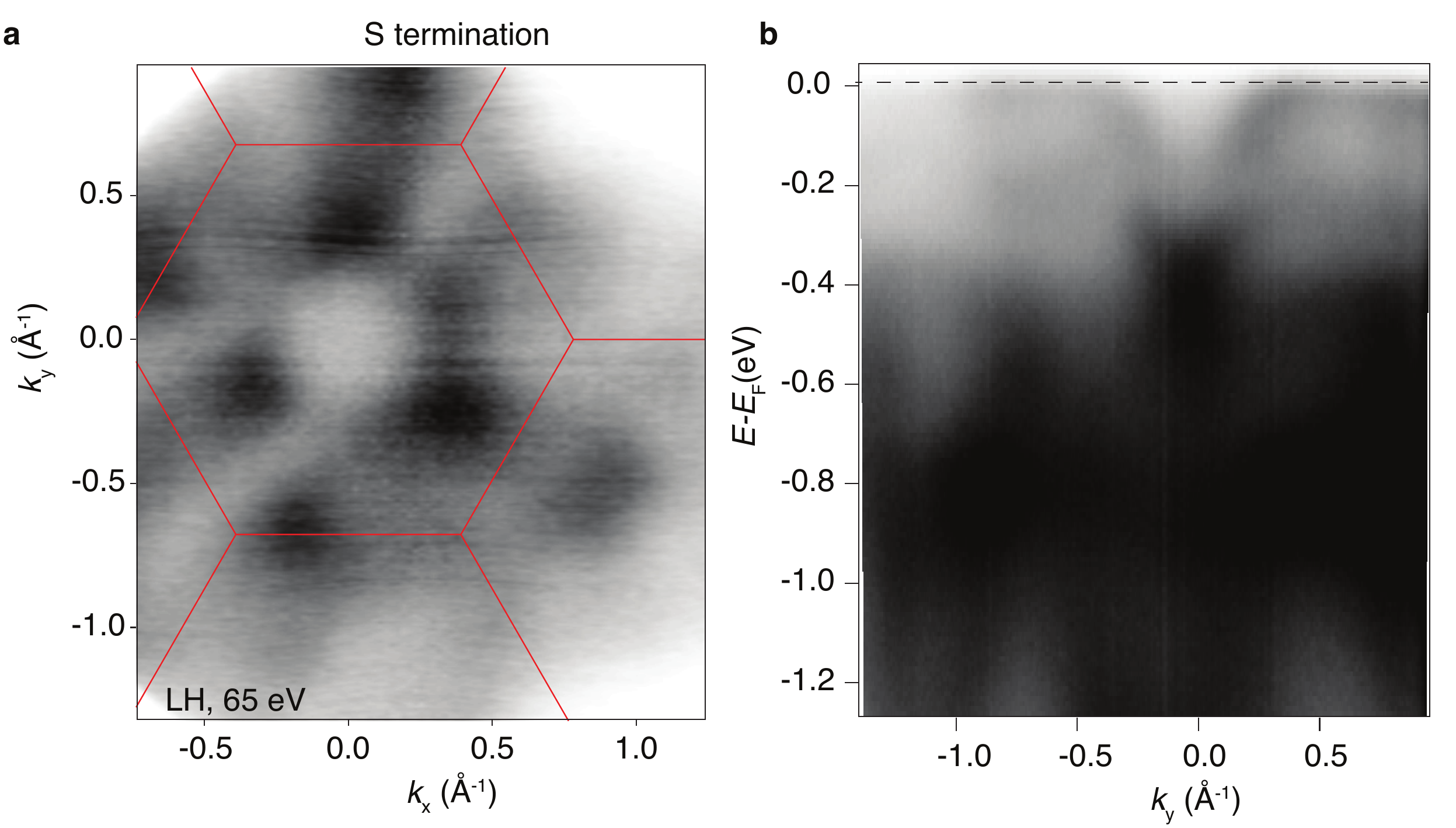}
\caption{(a) S termination Fermi surface collected with linear horizontal polarization and at 65 eV and its corresponding (b) energy-momentum dispersion along $\Gamma$-M. The high resolution data are in perfect agreement with those obtained by nano-ARPES and provide a complementary view of the states belonging to the Sn surface manifold.}
\label{S3}
\end{figure*}

\begin{figure*}
\centering
\includegraphics[width=\textwidth]{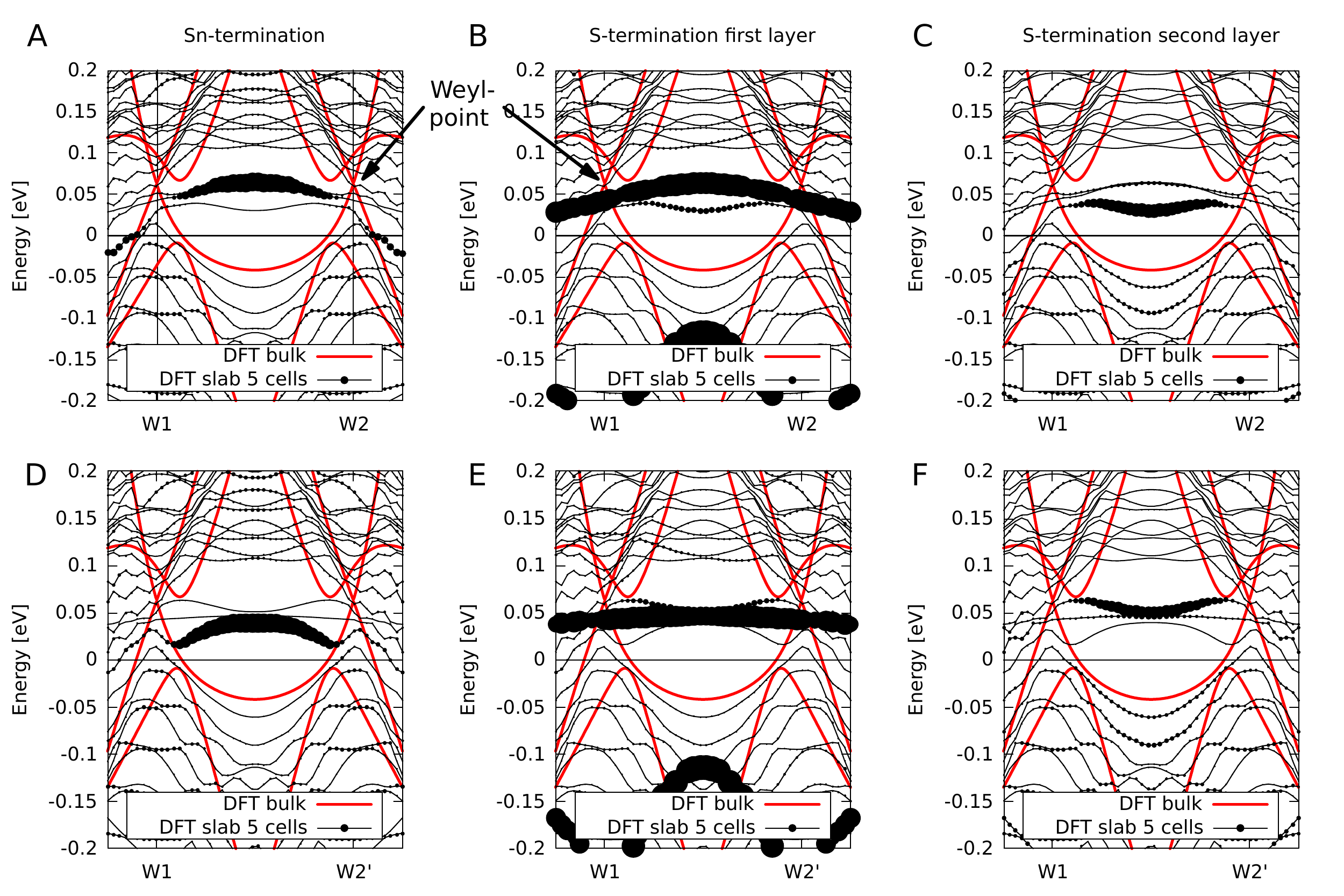}
\caption{(A-F) DFT slab band structure along different Weyl-point overlaid with the corresponding bulk band structure. The point size corresponds to the projection on the Sn-terminated surface and on the S-terminated surface for the first and second Co-kagome layer individually. For (A) and (F) non-trivial surface states can be identified, approximately connecting the Weyl-points at 60\,meV indicated by W1, W2 and W2'. W2 and W2' represent the two neighbouring Weyl-point in the Brillouin zone of Weyl-point W1. See Fig. \ref{S12}D.}
\label{S11}
\end{figure*}

\begin{figure*}
\centering
\includegraphics[width=\textwidth]{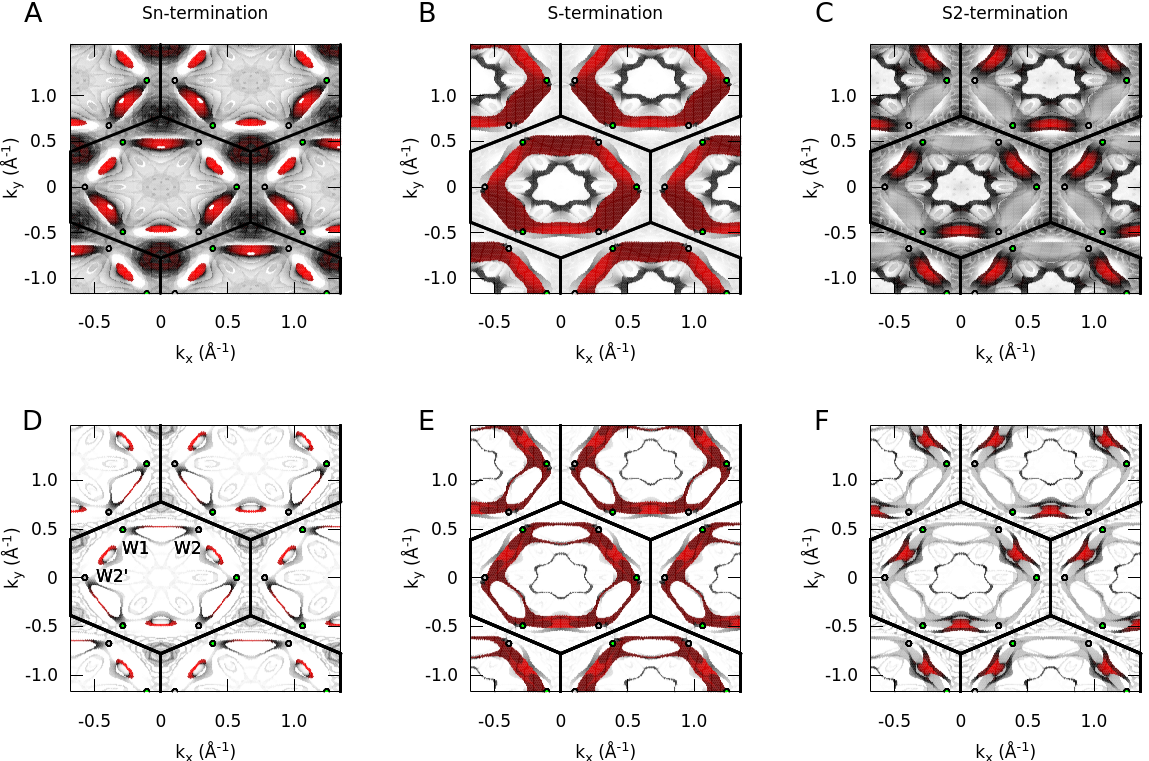}
\caption{(A-F) DFT Fermi surface maps for different terminations and energy integration ranges. For (A-C) a energy window of 30\,meV was used around 70\,meV above the Fermi level to capture the surface band around the Weyl-energy.  In (D-E) the energy range was reduced to 10\,meV around 50\,meV in order to show the typical Fermi arc representations in D and F. The states most dominant in B and E are trivial surface states. The surface character is indicated for increasing intensity from white, black to red color.}
\label{S12}
\end{figure*}

\end{document}